# SOCIAL MEDIA ANALYSIS FOR ORGANIZATIONS: US NORTHEASTERN PUBLIC AND STATE LIBRARIES CASE STUDY

**Matthew Collins**
University of South Carolina
mrc1@email.sc.edu

**Amir Karami**
University of South Carolina
karami@sc.edu

**ABSTRACT**

Social networking sites such as Twitter have provided a great opportunity for organizations such as public libraries to disseminate information for public relations purposes. However, there is a need to analyze vast amounts of social media data. This study presents a computational approach to explore the content of tweets posted by nine public libraries in the northeastern United States of America. In December 2017, this study extracted more than 19,000 tweets from the Twitter accounts of seven state libraries and two urban public libraries. Computational methods were applied to collect the tweets and discover meaningful themes. This paper shows how the libraries have used Twitter to represent their services and provides a starting point for different organizations to evaluate the themes of their public tweets.

**INTRODUCTION**

Public libraries engage their patrons and communities through several new participation platforms, such as social media and makerspaces, which enable and mediate new forms of online user participation, engagement, access, and interactivity (Cavanagh, 2016). Twitter is one such participation space adopted by public libraries as a means to expand the scope of their services and to increase their community connectivity (Al-Daihani and AlAwadhi, 2015; Cavanagh, 2016). Since its launch in July 2006, Twitter has quickly become one of the most popular social networking platforms for users to update their followers, and to provide convenient and effective information dissemination (Del Bosque, Leif, & Skarl, 2012; Shulman, Yep, & Tomé, 2015). For scholars in fields such as linguistics, sociology, economics, health, and psychology who are looking for real-time conversational or text data, Twitter has provided a wealth of data for behavioral analyses, sentiment analyses, trend analyses, information dissemination, and health surveillance (Karami, A., Dahl, A. A., Turner-McGrievy, G., Kharrazi, H., and Shaw, G., 2018). Mining these data provides an instantaneous snapshot of the content of public library tweets (Chew and Eysenbach, 2010). Social media sheds light on how patrons use library services, as well as assists academic libraries in advancing research (Yep, J., Brown, M., Fagliarone, G., and Shulman, J., 2017).

Beyond facilitating discussion between libraries, social media analysis (SMA) can also help libraries and other non-profit institutions to measure their costs and the satisfaction level of their users. When using SMA to their advantage, organizations are able to tailor their products to the needs of their clients & patrons. This is especially affordable for small and mid-size libraries that do not have the financial resources to afford traditional marketing campaigns. In addition to lower costs, libraries improve on serving their patrons when they utilize social media.

Twitter data analysis provides an additional convenient means of finding data for an organic user study. Twitter data have been applied in different domains such as health (Shaw, G. and Karami, A., 2017) and election analysis (Karami, A., Bennett, L. S., and He, X., 2018). Research has shown that when a library has an engaged and active Twitter following, it is able to spread information more easily (Yep, J., Brown, M., Fagliarone, G., and Shulman, J., 2017). Twitter shows both sides of user evolution; as users offer their opinions and needs, libraries have real-time feedback they can utilize to adjust their services. Simultaneously, Twitter gives libraries unprecedented access to their users. Social media in general allows libraries to engage users who may otherwise never think of what the library can do for them. In addition to proactive community outreach programs, Twitter enables libraries to advertise their services and programs to many more users than traditional media. In this study, we mine and analyze public library-related topics extracted from the Twitter accounts of nine public libraries in the northeastern United States of America in order to examine how public libraries use social media to describe their services and interact with patrons. The goal of this research is to find semantic patterns in thousands of tweets with respect to different libraries.

**RELATED WORKS**

Due to its popularity as an approach to analyze and understand large datasets not amenable to traditional qualitative research techniques, text mining tools have been applied to a range of information problems, such as understanding themes in social media or facilitating information retrieval in unstructured data (Maceli, 2016). Several previous studies have used text mining techniques to analyze tweet content from diverse fields, such as mental health (Jamison-Powell et al., 2012), academic libraries (Stvilia & Gibradze, 2017), and journalism and mass communication (Guo et al., 2016). One study applied text mining on unstructured text content to search for patterns in the tweets posted by the three largest American pizza chains: Papa John's, Pizza Hut, and Domino's Pizza. Their results reveal the value of text mining as an effective technique to extract business value from tweet content (He, Zha, & Li, 2013). According to the State of America's Libraries Report 2014, 84% of the largest libraries in the USA have Twitter accounts (Del Bosque, Leif, & Skarl, 2012; ALA, 2014). Many studies have examined the content of library tweets, as well as the effectiveness of Twitter usage by libraries (Al-Daihani & AlAwadhi, 2015; Yep et al., 2017). This study explored tweet content and the use of Twitter in 15 public libraries and 15 academic libraries in order to understand the libraries' Twitter-use patterns. The analysis emphasized three findings: (1) there was no significant difference in the frequency that public and academic libraries used Twitter, (2) academic libraries used formal language more frequently than public libraries, and (3) while there were content differences between public and academic library tweets, Twitter enabled both kinds of libraries to broadcast and share information about their activities, opinions, status, and professional interests (Aharony, 2010).

The huge amount of text data in different applications has created a need for text mining to detect interesting pattern. Text mining has been used for different applications such as spam detection (Karami A. and Zhou L., 2014; Karami, A. and Zhou, B., 2015), and health corpora analysis (Karami, A., Gangopadhyay, A., Zhou, B., and Kharrazi, H., 2015a). Among different text mining methods, topic modeling, especially Latent Dirichlet Allocation (LDA) model, is a successful computational approach for disclosing hidden topics in a Twitter corpus (Karami, A., 2015; Guo et al., 2016). Some studies have manually examined the content of public library or academic library tweets as an exploration of Twitter use in libraries (Aharony, 2010). However, to our knowledge no other study has used computational methods to analyze tweet content posted by public libraries in the United States. Our study applied LDA to analyze the contents of tweets posted by public state libraries.

**METHODOLOGY & RESULTS**

**Data Collection**

For this study, we collected data from the Twitter accounts of seven state public libraries and two public libraries in the northeastern United States of America using the Twitter API using the twitteR package in the R platform. The tweets were extracted from the Twitter accounts of the *Massachusetts State Library, the Connecticut State Library, the Maine State Library, the New Hampshire State Library, the New Jersey State Library, the Office of Library & Information Services of Rhode Island*, and the *Vermont Department of Libraries*, as well as from the Twitter accounts of the *New York Public Library* and the *Free Library of Philadelphia*. *The New York Public Library* and *Free Library Twitter* accounts were used because the state libraries of New York and Pennsylvania did not have twitter accounts. In total, 19,199 tweets were extracted in December 2017.

**Computational Content Analysis**

Semantic analysis of thousands of tweets is not an easy task, nor an efficient strategy to manually explore the contents of tweets; therefore, there is a need to apply computational methods to extract hidden topics in a corpus. Therefore, we applied topic modeling that is a machine learning process that uses algorithms to find out themes or topics prominent within documents in a large collection of documents. Fundamentally, a topic model automatically identifies topics within a collection of documents based on the words contained in each document (Cain, 2017; Gosh & Guha, 2014). This study implements a frequently used LDA topic model.

In topic analysis studies of Twitter posts, LDA has been most commonly applied to the analysis of well-constructed text documents, such as newspaper and academic journal articles, which are reviewed and edited (Kim et al., 2016; Karami, A., Gangopadhyay, A., Zhou, B., & Karrazi, H., 2015b). Fundamentally, the LDA model repeatedly samples the words from a collection of tweets to identify which words tend to associate with each other. The result of this sampling process is that words are assigned to multiple topics with differing probabilities (Gosh & Guha, 2014). LDA method assigns the related words into different groups, wherein each group represents a theme. For example, LDA assigns "Human Resource", "Marketing", "Finance", and "Business" into a group whose theme (label) is "Business" (Karami, A., Gangopadhyay, A., Zhou, B., and Kharrazi, H., 2017).

**Topic Discovery and Analysis**

Topical analysis included four steps: 1) labeling the topics, 2) detecting the frequency of the topics for each of the states, 3) calculating the weights of topics, and 4) recognizing the interesting patterns. To determine topic labels, we utilized categories and sub-categories created from previous studies (Aharony, 2010; Cavanagh, 2016). In the first step, we chose two categories for 10-word topic groups: 'library in general' and 'information about'. Sub-categories in the 'library in general' category included 12 labels. Topics in the 'library in general' category refer to tweet themes specifically related to the library. Sub-categories in the 'information about' category included seven labels. Topics in the 'information about' category refer to themes that were miscellaneous or not related to the library. For the libraries in the dataset, the most common sub-categories were library services, library events, book recommendations, library collection, library in general, and theme days and months. For the libraries in the dataset, the least common sub-categories were references, links, book reviews, author information, blogs, and professional articles (Tables 1 – 4).

These four tables provide the distribution and description of the detected topics. Table 2 shows that the library events, library collection, and library services were the most common sub-topics, and the blogs and the article professional topics were the uncommon topics for library in general topics. This table also indicates that the Twitter account of the Vermont library covering 8/12 topics, and the Main and the New Hampshire libraries covering 5/12 topics represented the highest and the lowest diversity of the topics, respectively. On other hand, it is interesting that the non-library events and the link were the most common and uncommon topics, respectively in Table 4. The library in Massachusetts covered most of topics and the libraries in New Jersey and New York covered just one topic with respect to "Information About" category in Twitter.

**CONCLUSION**

Social media growth has encouraged organizations to make a new communication channel for presenting their products and services. Millions of online comments post every day that it is impossible to analyze them manually. This paper proposes a computational approach to collect and analyze millions of tweets. In this study, we analyzed a corpus of tweets using topic modeling to examine how public libraries can use Twitter to describe their services and interact with patrons. After examining our prescribed topic labels, the results indicated that most tweets were related to library services, events, programs, and resources. These findings show that public libraries primarily use their Twitter accounts to communicate with their patrons and community. Choosing a set of topics provided a broad scope to generalize on themes mentioned by not one specific library, but several libraries. In addition, our computational approach removes the bias in the human component of manual content analysis.

This research opens a new direction for analyzing a huge amount of data in social media with respect to any organizations. This paper's approach can help to compare social media strategy of the organizations and detect success factors. The findings in this study can also serve as a springboard for future Twitter-based research, such as analyzing and categorizing the libraries in other states. Text mining could also be used to observe and track the growth of libraries.

| Table 1: Library in General Topic Descriptions | | |
|---|---|---|
| Topic Number | Associated Words for Topic / Topic Description | Topic Label |
| 1 | Associated words: ticket, events, join, discuss, location<br>Describing an event scheduled to occur at a library. | Library events |
| 2 | Associated words: books, list, best, favorite, recommendations<br>Describing books recommended by library staff for patrons. | Book recommendations |
| 3 | Associated words: collections, new, digital, audiobooks, photos<br>Describing items included in library collection. | Library collection |
| 4 | Associated words: free, skills, summer, reading, resources<br>Describing services and resources provided by the library. | Library services |
| 5 | Associated words: reference, web, sites, index, information<br>Describing reference resources available in library. | References |
| 6 | Associated words: librarians, funding, public, library, building, card,<br>Describing General information concerning the library. | Library in general |
| 7 | Associated words: book, good, great, read, pick<br>Reviewing of the quality, author, plot, subject of a book. | Book reviews |
| 8 | Associated words: women, year, books, history, week<br>Describing upcoming celebration to promote a program, issue, or campaign. | Theme days and months |
| 9 | Associated words: new, authors, birthday, happy, author<br>Describing information regarding an author. | Author information |
| 10 | Associated words: blog, read, post, check, latest, news<br>Describing blogs recommended by library staff. | Blogs |
| 11 | Associated words: conference, public, library, register, books, ala<br>Describing information related to upcoming professional or local conferences. | Conferences |
| 12 | Associated words: library, journal, article, press, newsletter<br>Describing notice of or links to library profession articles. | Professional Articles |

| Table 2: Distribution of 'Library in General' Topics | | | | | | | | | | | | | |
|---|---|---|---|---|---|---|---|---|---|---|---|---|---|
| States | Total Number of Tweets | Topic 1 | Topic 2 | Topic 3 | Topic 4 | Topic 5 | Topic 6 | Topic 7 | Topic 8 | Topic 9 | Topic 10 | Topic 11 | Topic 12 |
| Connecticut, CT | 3200 | 1 | 1 | 1 | 1 | 0 | 1 | 0 | 1 | 1 | 0 | 0 | 0 |
| Maine, ME | 3191 | 1 | 0 | 1 | 1 | 0 | 1 | 0 | 1 | 0 | 0 | 0 | 0 |
| Massachusetts, MA | 1293 | 1 | 1 | 1 | 1 | 0 | 0 | 0 | 1 | 0 | 1 | 0 | 0 |
| New Hampshire, NH | 393 | 1 | 0 | 1 | 1 | 0 | 1 | 0 | 1 | 0 | 0 | 0 | 0 |
| New Jersey, NJ | 3200 | 1 | 1 | 1 | 1 | 0 | 1 | 0 | 0 | 0 | 0 | 1 | 0 |
| NYPL, NY | 3200 | 1 | 1 | 1 | 1 | 0 | 1 | 0 | 1 | 1 | 0 | 0 | 0 |
| Free Library, PA | 3199 | 1 | 1 | 1 | 1 | 0 | 1 | 0 | 1 | 0 | 0 | 0 | 0 |
| Rhode Island, RI | 448 | 1 | 0 | 1 | 1 | 0 | 1 | 1 | 1 | 0 | 0 | 1 | 0 |
| Vermont, VT | 1075 | 1 | 1 | 1 | 1 | 0 | 1 | 0 | 1 | 0 | 0 | 1 | 1 |

| Table 3: 'Information about' Topic Descriptions | | |
|---|---|---|
| Topic Number | Associated Words for Topic / Topic Description | Topic Label |
| 1 | Associated words: thanks, people, share, books, lovely / Describing posts that express gratitude and appreciation. | Thanks |
| 2 | Associated words: congratulations, service, honor, prize, award / Describing achievements and accomplishments. | Congratulations |
| 3 | Associated words: health, health-connect, opioid, treatment, centers / Describing health-related topics. | Health |
| 4 | Associated words: event, park, museum, classic, celebrate / Describing upcoming events that will not take place at the library. | Non-library Events |
| 5 | Associated words: website, access, online, resources, election / Describing messages that include hyperlinks. | Links |
| 6 | Associated words: new, effective, today, senate, local, weather / Concerning community news and information. | Local News |
| 7 | Associated words: historical, society, places, photos, newspapers / Describing local history and historical facts. | Local History |

| Table 4: Distribution of 'Information about' Topics | | | | | | | |
|---|---|---|---|---|---|---|---|
| State | Topic 1 | Topic 2 | Topic 3 | Topic 4 | Topic 5 | Topic 6 | Topic 7 |
| Connecticut, CT | 1 | 0 | 0 | 1 | 0 | 0 | 1 |
| Maine, ME | 1 | 1 | 0 | 1 | 0 | 0 | 1 |
| Massachusetts, MA | 1 | 0 | 0 | 1 | 1 | 1 | 1 |
| New Hampshire, NH | 1 | 0 | 0 | 1 | 0 | 1 | 1 |
| New Jersey, NJ | 0 | 0 | 0 | 1 | 0 | 0 | 0 |
| NYPL, NY | 0 | 1 | 0 | 0 | 0 | 0 | 0 |
| Free Library, PA | 0 | 0 | 1 | 1 | 0 | 1 | 1 |
| Rhode Island, RI | 0 | 0 | 1 | 0 | 0 | 1 | 0 |
| Vermont, VT | 1 | 1 | 1 | 1 | 0 | 0 | 0 |

**REFERENCES**

1. Aharony, N. (2010). Twitter Use in Libraries : An Exploratory Analysis. *Journal of Web Librarianship*, 4(4), 333–350. https://doi.org/10.1080/19322909.2010.487766

2. Al-Daihani, S. M., & AlAwadhi, S. A. (2015). Exploring academic libraries ' use of Twitter : a content analysis. *The Electronic Library*, 33(6), 1002–1015. https://doi.org/10.1108/EL-05-2014-0084

3. American Libraries Association (ALA). (2014). State of America's Libraries Report 2014. Retrieved from http://www.ala.org/news/state-americas-libraries-report-2014

4. American Libraries Association (ALA). (2017). State of America's Libraries Report 2017. Retrieved from http://www.ala.org/news/state-americas-libraries-report-2017

5. Cain, J.O. (2017). Using Topic Modeling to Enhance Access to Library Digital Collections. *Journal of Web Librarianship*, (10)3, 210–225. https://doi.org/10.1080/19322909.2016.1193455

6. Cavanagh, M. F. (2016). Micro-blogging practices in Canadian public libraries : A national snapshot. Journal of Librarianship and Information Science, 48(3), 247–259. https://doi.org/10.1177/0961000614566339
7. Chew, C., & Eysenbach, G. (2010). Pandemics in the Age of Twitter : Content Analysis of Tweets during the 2009 H1N1 Outbreak. PLoS ONE, 5(11), 1–13. https://doi.org/10.1371/journal.pone.0014118
8. Del Bosque, D., Leif, S. A., & Skarl, S. (2012). Libraries atwitter: trends in academic library tweeting. Reference Services Review, 40(2), 199–213. https://doi.org/10.1108/00907321211228246
9. Ghosh, D. (Debs), & Guha, R. (2014). What are we "tweeting" about obesity? Mapping tweets with Topic Modeling and Geographic Information System. Cartography and Geographic Information Science, (40)2, 90–102. https://doi.org/10.1080/15230406.2013.776210
10. Guo, L., Vargo, C. J., Pan, Z., Ding, W., & Ishwar, P. (2016). Big Social Data Analytics in Journalism and Mass Communication : Comparing Dictionary-Based Text Analysis and Unsupervised Topic Modeling. Journalism & Mass Communication Quarterly, 93(2), 332–359. https://doi.org/10.1177/1077699016639231
11. He, W., Zha, S., & Li, L. (2013). Social media competitive analysis and text mining : A case study in the pizza industry. International Journal of Information Management, 33, 464–472. https://doi.org/10.1016/j.ijinfomgt.2013.01.001
12. Jamison-Powell, S., Linehan, C., Daley, L., Garbett, A., & Lawson, S. (2012). "I can't get no sleep": Discussing #insomnia on Twitter. In Proceedings of the SIGCHI Conference on Human Factors in Computing Systems (pp. 1501–1510). Retrieved from http://andygarbett.co.uk/wp-content/uploads/2016/05/Jamison-Powell-et-al-I-Cant-Get-No-Sleep.pdf
13. Karami A. and Zhou L. (2014), Improving Static SMS Spam Detection by Using New Content-based Features, Proceedings of the 20th Americas Conference on Information Systems (AMCIS), Savannah, GA.
14. Karami, A. (2015). Fuzzy topic modeling for medical corpora. University of Maryland, Baltimore County.
15. Karami, A., Gangopadhyay, A., Zhou, B., and Kharrazi, H. (2015a). A fuzzy approach model for uncovering hidden latent semantic structure in medical text collections. iConference 2015 Proceedings.
16. Karami, A., Gangopadhyay, A., Zhou, B., and Karrazi, H. (2015b). Flatm: A fuzzy logic approach topic model for medical documents. In 2015 Annual Conference of the North American Fuzzy Information Processing Society (NAFIPS) held jointly with 2015 5th World Conference on Soft Computing (WConSC), (pp. 1-6). IEEE.
17. Karami, A. and Zhou, B. (2015). Online review spam detection by new linguistic features. iConference 2015 Proceedings.
18. Karami, A., Bennett, L. S., and He, X. (2018). Mining Public Opinion about Economic Issues: Twitter and the US Presidential Election. International Journal of Strategic Decision Sciences (IJSDS), 9(1), 18-28.
19. Karami, A., Dahl, A. A., Turner-McGrievy, G., Kharrazi, H., and Shaw, G. (2018). Characterizing diabetes, diet, exercise, and obesity comments on Twitter. International Journal of Information Management, 38, 1, 1-6.
20. Karami, A., Gangopadhyay, A., Zhou, B., and Kharrazi, H. (2017). Fuzzy approach topic discovery in health and medical corpora. International Journal of Fuzzy Systems, 1-12.
21. Kim, E. H., Jeong, Y. K., Kim, Y., Kang, K. Y., & Song, M. (2016). Topic-based content and sentiment analysis of Ebola virus on Twitter and in the news. Journal of Information Science, 42(6), 763–781. https://doi.org/10.1177/1045389X14554132
22. Maceli, M. (2016). Introduction to Text Mining with R for Information Professionals. The Code4Lib Journal, (33), 1–7. Retrieved from http://journal.code4lib.org/articles/11626
23. Shaw, G., and Karami, A. (2017). Computational content analysis of negative tweets for obesity, diet, diabetes, and exercise. Proceedings of the Association for Information Science and Technology, 54(1), 357-365.
24. Shulman, J., Yep, J., & Tomé, D. (2015). Leveraging the Power of a Twitter Network for Library Promotion. The Journal of Academic Librarianship, 41(2), 178–185. https://doi.org/10.1016/j.acalib.2014.12.004
25. Yep, J., Brown, M., Fagliarone, G., & Shulman, J. (2017). Influential Players in Twitter Networks of Libraries at Primarily Undergraduate Institutions. The Journal of Academic Librarianship, 43(3), 193–200. https://doi.org/10.1016/j.acalib.2017.03.005